\documentstyle[12pt,aaspp4,flushrt]{article}

\begin{document}


\title{An Exploration of the Paradigm for the 2-3 Hour Period Gap in
Cataclysmic Variables}

\author{Steve B. Howell}
\affil{Astrophysics Group, Planetary Science Institute \\
620 N. 6th Avenue, Tucson, AZ  85705}

\author{Lorne A. Nelson$^1$}
\affil{Canadian Institute for Theoretical Astrophysics \\
University of Toronto, 60 St. George Street, Toronto, Ontario, Canada M5S 3H8}

\and

\author{Saul Rappaport}
\affil{Department of Physics and Center for Space Research \\
Massachusetts Institute of Technology, Cambridge, MA 02139}

\begin{center}
\sl{The Astrophysical Journal}, in press
\end{center}

$^1$Present address: Physics Dept., Bishop's University, Lennoxville, QC, \\
    Canada J1M 1Z7

\begin{abstract}
We critically examine the basic paradigm for the origin of the
2-3 hr period gap in cataclysmic variables (CVs), i.e., binary
systems in which a white dwarf accretes from a relatively unevolved,
low-mass donor star.   The observed orbital period distribution for
$\sim300$ CVs shows that these systems typically have orbital periods,
$P_{orb}$,
in the range of $\sim80$ min to $\sim8$ hr, but a distinct dearth of systems
with 2 $\la P_{orb}$(hr) $\la 3$. This latter feature of the period
distribution is often referred to as the ``period gap".  The conventional
explanation for the
period gap involves a thermal bloating of the donor star for $P_{orb} \ga 3$
hr due to mass transfer rates which are enhanced over those which
could be driven by gravitational radiation (GR) losses alone (e.g.,
magnetic braking).  If for some reason the supplemental
angular momentum losses become substantially reduced when $P_{orb}$
decreases below $\sim3$ hr, the donor star will relax thermally and shrink
inside of its Roche lobe.  This leads to a cessation of mass transfer until GR
losses can bring the system into Roche-lobe contact again at $P_{orb} \sim
2$ hr.

We carry out an extensive population synthesis study of CVs
starting from $\sim 3 \times 10^6$ primordial binaries, and
evolving some $\sim 2 \times 10^4$ surviving systems through their CV
phase.  In particular we study current-epoch distributions of CVs in the
$\dot M-P_{orb}$, $R_{2}-P_{orb}$, $M_{2}-P_{orb}$, $q-P_{orb}$,
$T_{eff}-P_{orb}$, and $L_{2}-P_{orb}$ planes,
where $\dot M$ is the mass transfer rate, $q$ is the mass ratio $M_2$/$M_1$,
and $M_2$, $R_2$,
$T_{eff}$, and $L_2$ are the donor star mass, radius, effective
temperature, and luminosity, respectively.  This work presents a new
perspective on theoretical studies of the long-term evolution of CVs.
In particular, we show that if the current paradigm is correct, the
secondary masses in CVs just above the period gap should be as
much as $\sim 50\%$ lower than would be inferred if one assumes
a main-sequence radius-mass relation for the donor star.
We quantify the $M_{2}-P_{orb}$ relations expected
from models wherein the donor stars are thermally bloated.  Finally,
we propose specific observations, involving the determination of
secondary masses in CVs, that would allow for a definitive test
of the currently accepted model (i.e., interrupted thermal bloating)
for the period gap in CVs.
\end{abstract}

{\it Subject headings:} cataclysmic variables
$-$ stars: binaries: close
$-$ stars: evolution
$-$ stars: mass loss
$-$ stars: low mass

\newpage

\section{Introduction}

Cataclysmic variables (CV) are short period binary systems
consisting of a white dwarf that accretes matter via Roche-lobe
overflow from a low-mass companion star.  These objects
exhibit a wide range of phenomenology including optical
flickering in nova-like systems, dwarf nova eruptions
which are thought to be caused by thermal instabilities
in the accretion disks, and classical nova explosions which
are thermonuclear runaways of the accreted matter on the
white dwarf (see, e.g., Warner 1995).  The range of
observed phenomena depends on the mass transfer
rate, the mass ratio of the stellar components, and the magnetic
field strength of the accreting white dwarf.  The orbital periods of
the majority of CVs range from 8 hours down to about 78 minutes,
but both longer and shorter period systems are known.
In the former case, the donor stars are typically somewhat evolved,
while in the latter case, the donor stars are hydrogen exhausted.
In this paper we focus on the gap that exists
in the orbital period distribution of CVs in the range of
$\sim 2-3$ hr (see, e.g., Warner 1976; Rappaport, Verbunt, \& Joss 1983,
hereafter RVJ; Spruit \& Ritter 1983; Hameury, et al.
1988a; Warner 1995).

The overall evolution of CV binaries is thought to be fairly
well understood. The widely accepted explanation for the period
gap rests on a mechanism for extracting angular momentum from
the binary orbit (e.g., via magnetic braking of the secondary)
for periods down to $\sim3$ hours, followed by a relatively
substantial decrease in the angular momentum loss rate\footnote{
We note that this scenario does not require the angular momentum
loss rate to drop suddenly. Instead, it requires only that the
timescale over which the angular momentum loss rate decreases must be shorter
than the
thermal timescale of the donor.}.
The donor star, which had been thermally
``bloated'' in response to the mass loss driven by the systemic
angular momentum losses, is then able to relax inside of its
Roche lobe and mass transfer ceases.  The donor star is
then thought to reestablish Roche-lobe contact by the time
the orbital period has decreased to about 2 hr,
after which mass transfer resumes.   In this paper we
critically examine this paradigm for the creation of the
period gap.  While most workers believe in the existence of the so-called
``2-3 hr period gap", a few (e.g., Wickramasinghe \& Wu 1994; Verbunt
1997, but see also Warner 1995; Wheatley 1995)
have questioned its reality, especially when all types of CV
are considered; however, we adopt the view that the period
gap is a real feature of the CV population as a whole and, as such,
requires a theoretical explanation (with observational tests)
within the context of their binary evolution. Finally in this regard
we note a suggestion by Clemens et al. (1998) that
the period gap results from a ``kink'' in the radius-mass
relation for main-sequence stars at a mass of about
$\sim 0.25~M_\odot$ (but see the rebuttal by Kolb, King, \&
Ritter 1998).

In \S 2 we describe the conventional picture of
the evolution of a typical CV, including the period gap,
and show some illustrative examples of binary evolution
calculations for individual systems. In \S 3 we explore
how the binary evolution alters the relations among mass, radius,
and orbital period of the secondary star.  Specifically we discuss
how the main-sequence radius-mass relationship must be modified to
include the addition of a ``bloating factor'' that accounts for the changes
caused by departures from thermal equilibrium
of the mass-losing secondary star. We derive semi-analytic
mass-period and radius-period relationships
for CV secondaries. In \S 4 we describe our population
synthesis and binary evolution codes, while in \S 5 we
present results from our population synthesis study of CVs in
which the binary parameters of the CVs at all phases
of their evolution are explored.  In \S 6 we show how assumptions
that the donor star has a main-sequence radius-mass relation
can lead to large errors in the assignment of the constituent stellar
masses, most notably within the orbital period range of 3-5 hr.
This period range should encompass the maximum bloating exhibited
by a CV secondary compared to a main-sequence star of the same mass.
Also in \S 6 we discuss some specific observational implications
resulting from our theoretical work.  In particular, a specific test
for CVs just above the period gap which will enable us, in principle,
 to distinguish unambiguously among different possible explanations
for the period gap is presented.  Finally, we present our summary
and conclusions in \S 7.

\section{Standard Evolutionary Scenario for CVs}

In the conventional picture of CV evolution (see, e.g.,
Faulkner 1971; Paczy\'nski \& Sienkiewicz 1981; Rappaport, Joss, \&
Webbink 1982, hereafter RJW; RVJ; Spruit \& Ritter 1983;
Hameury et al. 1988a; Kolb 1993), the early phases are expected
to be dominated by angular momentum losses due to magnetic
braking via a magnetically constrained stellar wind from
the donor star (see, e.g., Verbunt \& Zwaan 1981; RVJ).
In these early phases, mass transfer rates are typically $\sim
10^{-9}$ to $ 10^{-8} M_\odot~{\rm yr^{-1}}$, and orbital
periods range from $\sim$ 8~hr to $\sim$ 3 hr, just at the
upper edge of the period gap.  At some point in the evolution,
the secondary becomes completely convective (at $\sim 0.23~M_\odot$)
and, in the currently accepted view, magnetic braking is
assumed to be greatly reduced.  The near cessation of
magnetic braking reduces the mass transfer rate, and
allows the secondary to shrink toward its thermal equilibrium
radius.  This causes a period of detachment, during which $\dot M$
drops to essentially zero, and which lasts until the Roche lobe shrinks
sufficiently to bring the secondary back into contact with it,
at an orbital period of $\sim 2$ hr.  This is the commonly
accepted explanation for the observed period gap between
2-3 hrs in CVs (RVJ; Spruit \& Ritter 1983).

When mass transfer recommences at $P_{orb}$ $\sim$ 2 hrs,
it is then driven largely by gravitational radiation
losses at rates of $\sim 10^{-10} M_\odot~{\rm yr^{-1}}$.
As the orbit shrinks and the mass of the donor star
decreases, the mass-loss timescale increases, but the thermal
timescale, $\tau_{KH}$, increases much faster, due to its approximate
$\sim M^{-2}$ dependence. Therefore, at some point the thermal
timescale grows larger than the mass-transfer timescale.
When this occurs, the donor star is unable to
adjust to the mass loss on its thermal timescale, and it therefore
starts to expand upon further mass loss, in accordance with its
adiabatic response; i.e., $[{\rm{dln(R)/dln(M)}}]_{ad} < 0$.
The orbital period at this point is typically $\sim 80$ min
and the mass of the donor star is $\sim 0.06~M_\odot$. From
this point on, the mass of the donor star will continue to decrease
(but with longer and longer $\dot M$ timescales),
the orbital period will increase back toward periods approaching
$\sim$2 hr (within a Hubble time), and the interior of the
donor star will become increasingly electron degenerate.
A discussion of this later stage of CV evolution is presented by
Howell, Rappaport, \& Politano 1997; hereafter HRP).

To make these evolutionary descriptions somewhat more
quantitative, we show in Figure 1 the secular evolution
of several model CVs under the influence of magnetic braking
and gravitational radiation.  The evolution code used to
generate these results is a descendant of the one used by
RVJ, and is described in \S 4.2 along with recent
improvements to the code. The two panels on the left side of Fig. 1 show
the evolution with time of a CV binary with initial constituent masses of
$M_2  = 0.9~M_{\odot}$ and $M_{WD} = 1.1~M_{\odot}$, where
$M_2$ and $M_{WD}$ are the masses of the donor star and white dwarf,
respectively.  Other parameters used in the calculation are
for our ``Standard Model" (see Table 1 for definitions).
The top and bottom panels show the evolution
with time of the mass transfer rate and orbital period,
respectively, for an assumed donor star with solar composition.  The
calculations have been carried out to approximately the age of the Galaxy.
The evolutionary phases and features discussed above are present in Fig. 1,
including the interval where mass transfer is driven by magnetic braking
($\sim 10^{7.3} - 10^{8.4}$ yr), the period gap ($\sim 10^{8.4} -
10^{8.8}$ yr), the interval where $\dot M$ is driven by gravitational radiation
losses ($> 10^{8.8}$ yr), the period minimum at $10^{9.4}$ yr, the
subsequent increase in $P_{orb}$ back up to $\sim 2$ hrs, and the sharp
falloff in $\dot M$ after orbital period minimum.

On the right side of Fig. 1 the temporal evolution of four other
illustrative model CV binaries are shown.  The following discussion
contains descriptions of the period gap which develops in these systems;
these are easier to visualize by looking also at Fig. 2.  The initial masses
($M_{2}, M_{WD}$) of these systems are (0.2, 0.4), (0.35, 0.35),
(0.3, 0.6), and (0.65, 0.7), all in
units of $M_\odot$. For the system with initial masses (0.2, 0.4) the
binary comes into Roche-lobe contact for the first time at an orbital
period below the gap, i.e., at $P_{orb} = 2$ hr (solid curve in Figs.
1 c and d).  Note the enhanced mass transfer rate at $\sim$ 30 Myr after
Roche-lobe contact is made.  The subsequent evolution is not dissimilar to
the one shown in the left panels.  For the system with initial masses
(0.35, 0.35), the donor star commences mass transfer at a period of 3.3 hr,
with magnetic braking still operative (dotted curve in Figs. 1 c and d).
Because the two masses are the same when the donor star first fills
its Roche lobe, the mass transfer is only marginally stable
(see the discussion in \S 4.2 below).  Therefore, $\dot M$ is initially very
high and the system is quickly driven out of thermal equilibrium, causing
the orbit to expand.  This system comes out of contact (i.e., enters the
period gap) at an orbital period of 3.5 hr.  The system with initial masses
(0.3, 0.6) is an example of one that commences mass transfer {\it in} the
period gap.  Lastly, the system with initial masses (0.65, 0.7) is another
example of a system which exhibits the ``usual" 2-3 hr period gap, but
commences mass transfer at $P_{orb} = 5$ hr.

In Figure 2 the same evolutions shown in Fig. 1 are again presented, but
this time the binary parameters are displayed as a function of the evolving
orbital period.  As in Fig. 1, the left panels are for initial masses ($M_{2},
M_{WD}$) of (0.9, 1.1), while the right panels are for initial masses of
(0.2, 0.4), (0.35, 0.35), (0.3, 0.6), and (0.65, 0.7).  The top, middle,
and bottom panels show the evolution of $\dot M$, $M_{2}$, and
$R_{2}$, respectively.  As mentioned above, the period gap is more evident
in Fig. 2 than it is in Fig. 1. We note here several unique features
associated with the evolution of individual CVs; an understanding
of these features will aid our interpretation of the results obtained
for an entire population of evolving CV systems (see \S 5). For
example, $\dot M$ typically exhibits a sharp spike at the onset of
mass transfer, (see also RVJ and Hameury et al. 1988b); this behavior will
appear in all of the two-dimensional ``images" we produce from the
population synthesis calculations in \S5.
The mass of the donor stays constant during its evolution through the period
gap since there is no mass transfer taking place at that
time - this is indicated by the horizontal lines in the middle panels.  The
abrupt shift in location between the $M_{2}-P_{orb}$ track above the
period gap and below the gap will be dramatically apparent in the
population synthesis results, and will have important consequences
that are discussed below.  Finally, the radius of the donor star decreases
sharply after the system enters the period gap; in fact, it is the shrinking
of the donor inside of its Roche lobe when the magnetic braking ceases
that is the putative cause of the period gap.  Again, the abrupt shift
between the $R_{2}-P_{orb}$ track above and below the period gap
will be very pronounced in the population synthesis results.

A noteworthy feature of Figures 1a, 1c, 2a, and 2d mentioned above is the sharp
rise in $\dot M$ whenever mass transfer has just commenced, including the first
time that the donor star fills its Roche lobe, and after the resumption of mass
transfer below the period gap.  This results from the fact that when a low-mass
star (i.e., $\la 0.5 M_\odot$) is in thermal equilibrium (i.e., the nuclear
luminosity, $L_{nuc}$, equals the bolometric luminosity, $L_{opt}$),
the sudden onset of mass transfer forces the star to expand because its
adiabatic index is negative (discussed above).  This expansion can cause
a temporarily anomalously high rate of mass transfer, viz, the episodes
of high $\dot M$ seen in Figures 1a, 1c, 2a, and 2d.  However, as
soon as the donor star expands, its core temperature drops slightly, and
$L_{nuc}$, which is a highly sensitive function of temperature, drops
dramatically.  This leads to a luminosity deficit wherein $L_{nuc} < L_{opt}$.
The star can then lose a net amount of energy, shrink, and approach
its new thermal equilibrium radius (appropriate to its lower mass)
on a Kelvin-Helmholtz (i.e., thermal) timescale.  During the mass
loss process, true thermal equilibrium is never reached, and the
luminosity deficit attains a value which is adequate to allow the
star to shrink continuously.  The above discussion explains the
transient episodes of higher transfer rates at the start of mass transfer
epochs, and the ``outlying" lower probability CV states we shall encounter
in \S 5. It also explains the thermal ``bloating" of the donor star which
is discussed in \S 3, and which will play a key role in the
observational test we propose in \S 6. (For earlier discussions
of some of these basic effects, see RJW and Hameury et al. 1988a.)

The five individual evolutions shown in Figs. 1 and 2 serve
to illustrate the range of interesting possibilities for CVs which commence
mass transfer with different mass ratios.  The population synthesis study
described in \S 5, explores these various possibilities in a more systematic
and complete way.

\section{Quantitative Effects of Thermal Bloating of the Secondary Star}

We start with the assumption that during mass transfer in a CV the Roche
lobe of the donor star is located within its atmosphere, i.e., the donor star
is ``filling" its Roche lobe (see Howell et al. 2000).
We then take the Roche-lobe radius of the
secondary star to be given by the simple
expression of Kopal (1959):
\begin{eqnarray}
R_{2} \simeq 0.46~a\left(\frac{M_2}{M_2+M_{WD}}\right)^{1/3}   .
\end{eqnarray}
This can be combined with Kepler's 3rd law to yield the well-known
relationship among the mass, radius, and orbital period of the donor star:
\begin{eqnarray}
P_{orb}(M_{2}, R_{2}) \simeq 9~{M_2}^{-1/2}{R_2}^{3/2}  ,
\end{eqnarray}
where $M_{2}, R_{2}$ and $P_{orb}$ are expressed in units of $M_\odot$,
$R_\odot$, and hours, respectively.  If we now assume that the radius of
the donor star is some factor $f$ times the radius it would have if it were
a normal main-sequence star, we can write
\begin{eqnarray}
R_{2} = f a {M_2}^{b}   ,
\end{eqnarray}
where we approximate the radius-mass relation for stars on the lower
main sequence (i.e., G to M stars) by $R_{2} = a {M_2}^{b}$ where
$a$ and $b$ are constants, and we refer
to $f$ as the ``bloating factor".  This bloating factor $f$ is simply a
measure of how much larger the radius of a CV secondary is than that of a
single, main-sequence star of the same mass due to the departure from
thermal equilibrium.  We can now combine equations
(2) and (3) to derive relations for the mass and radius of CV secondaries
as a function of the binary orbital period.
\begin{eqnarray}
M_{2} \simeq 9^{-2/(3b-1)}{P_{orb}}^{2/(3b-1)} \left({af}\right)^{-3/(3b-1)}
\end{eqnarray}
\begin{eqnarray}
R_{2} \simeq 9^{-2b/(3b-1)}{P_{orb}}^{2b/(3b-1)}
\left({af}\right)^{-1/(3b-1)}   .
\end{eqnarray}
For the purposes of this exercise, we take $a$ = 0.85 and $b$ = 0.85,
which we find by fitting a power law to the main-sequence models of
Dorman, Nelson, \& Chau (1989; hereafter DNC).  With these values
for the constants $a$ and $b$, the above equations simplify to:
\begin{eqnarray}
M_{2}(P_{orb}) \simeq 0.08 f^{-1.95} P_{orb}^{1.3}
\end{eqnarray}
\begin{eqnarray}
R_{2}(P_{orb}) \simeq 0.10 f^{-0.65} P_{orb}^{1.1}   .
\end{eqnarray}
where, again, $M_2$ and $R_2$ are in solar units and $P_{orb}$ is in hours.
The conclusions drawn from these expressions are somewhat counterintuitive
in that, for a CV at a given orbital period, if the donor star is {\it
bloated}, the proper radius and mass that should be inferred from the orbital
period are {\it smaller} than the values that would be inferred if the star
were on the main sequence (see also Beuermann et al. 1998).
In \S 6 we derive polynomial fits for $M_2(P_{orb})$
and $R_2(P_{orb})$ from our population synthesis study; the analytic
expressions given by equations (6) and (7) serve mainly to demonstrate
how these quantities scale with the bloating factor $f$.

\section{Population Synthesis Study}

The individual binary evolution runs shown in Figs. 1 and 2 for
several different combinations of initial constituent masses are
instructive, but do not ({\it i}) adequately sample the full
range of possible initial masses, nor ({\it ii}) provide us with the
distributions of CV binary properties at the current
epoch.  We have therefore undertaken a population synthesis
study of CVs which consists of two parts.  In the first part, we
utilize a Monte Carlo approach to select a large number
($\sim 3 \times 10^6$) of primordial binaries, and follow
the evolution of these systems to see which ones undergo a common
envelope phase. In such events, the envelope of the giant star engulfs the
secondary, leading to a spiral-in episode which leaves the secondary in a
close orbit with a white dwarf (i.e., the core of the primary star; see, e.g.,
(Paczy\'nski 1976; Webbink 1979; see also \S 4.1 for details).
Primordial binaries which are too wide will not undergo any significant
mass transfer and will not lead to the formation of CV systems - the
evolution of such wide binaries is not followed in the present study.
Successful systems which emerge from the first part of our population synthesis
caculations
are those which do undergo a common envelope phase and yield a close binary
consisting of a white dwarf and low mass ($\lesssim~1~M_\odot$) companion.
The second part of the population synthesis considers those white-dwarf
main-sequence binaries for which systemic angular momentum
losses, or a modest amount of evolution by the normal
companion star, can initiate Roche-lobe contact within a Hubble
time.  Each of these systems is then evolved in detail through
the mass-transfer phase (CV phase) until the donor star has been
reduced to a negligible mass (typically 0.03 $M_\odot$).

A number of prior population synthesis studies of cataclysmic variables
have been carried out. These include work by Politano (1988; 1996);
de Kool (1992); Kolb (1993); Di Stefano \& Rappaport (1994, for CVs
in globular clusters); and HRP (emphasizing systems which evolve beyond
the orbital period minimum).   The current study has several new features
and advantages over the previous studies.   First, we compute
probability density functions in two parameters, e.g., $\dot M: P_{orb}$
and $M_{2}: P_{orb}$ (see \S 4.3).  This way of studying
and evaluating the results of population synthesis calculations has
a distinct advantage over producing distributions of a single
parameter.  For example, we are able to quantitatively evaluate
phases of the evolution that are short lived or represent unusual
evolutionary states, (e.g., whenever Roche-lobe contact has just
been established or when the initial binary mass ratio is near unity).
Other examples include the ability to discern the spread in $\dot M$ at
a given orbital period, the distinction between
systems with He and CO white dwarfs, and the pronounced
depression in secondary mass at a given orbital period (for systems
just above the period gap).  A second advantage is that our code for
evolving the donor stars was originally developed to evolve brown
dwarfs of very low mass and to very old ages.  The code has been
well ``calibrated" against other more sophisticated ones that have
been used for the purpose of evolving brown dwarfs (see \S 4.1).
Finally, our population synthesis code which is used to generate
the zero-age CVs that are input to the binary evolution code provides an
independent check on previous work, and tests the sensitivity
of our conclusions to various uncertainties in the physics, initial
conditions, and other input parameters.

Finally, we mention a possibly important limitation on
the study we present here, which also applies to most other prior
work in this area.  We have considered only donor stars with
$M_{2} \leq 1~M_\odot$, and do not allow for the chemical (nuclear)
evolution the donor.  The latter approximation is realistic if the donor
star commences mass transfer within $\sim$3$\times$10$^9$ years of the
common envelope event, or if the donor has a mass of $\la 0.7~
M_\odot$.  These conditions apply to most of the systems that
successfully evolve through the CV phase in our calculations.
Furthermore, we find that only $\sim$ 5\% of all the stable mass-transferring,
zero-age CVs in our population synthesis study have secondaries which
are older than 1/3 of their main-sequence lifetime prior to the start of
mass transfer. Theoretically,
there should indeed be some CVs which evolve from donor stars
that are initially more massive than $1~M_{\odot}$, and they should
be followed in future population synthesis studies.  For the
present study, we simply assume that such systems, with donors
whose initial mass exceeds 1 $M_{\odot}$, do not contribute
substantially to the CV population and, above all, would not affect
our conclusions concerning systems near the period gap.

In this regard, recent work by Beuermann et al. (1998) examines
the properties of the secondary star in CVs in an effort to determine
if they are indeed similar to normal main-sequence stars.  They show
that, in the spectral type-$P_{orb}$ plane, the $\sim 50$ CVs with measured
spectral types lie below the expected relation for main-sequence stars
(i.e., they are cooler at a fixed value of $P_{orb}$). Beuermann et al.
conclude that for systems with $P_{orb} < 6$ hr this effect could result from
mass loss (see \S5), but that for longer orbital periods this effect suggests
chemical evolution of the donor star.  This is a potentially important
finding for systems with orbital periods longer than we consider here,
and could also possibly impact the shorter period systems as well.
>From our population synthesis results, we find that $\sim10 \%$ of
CVs could potentially form with progenitors whose mass is initially
sufficiently high (i.e., $\gtrsim 1 M_\odot$) that chemical evolution of
the donor star might indeed be significant.  Systems with such donor stars
are not followed in the present study.  If, for some as yet unknown
reason, the more massive donor stars have a greater efficiency
for producing CVs than their lower-mass counterparts, then chemical
evolution may indeed prove influential in the evolution of CVs.  These
possibilities should be examined in future population synthesis studies.

\subsection{Choosing the Zero-Age CVs}

The properties of the primordial binary systems are chosen
via Monte Carlo techniques as follows.  The primary mass
is picked from Eggleton's (2000) Monte Carlo representation of the
Miller \& Scalo (1979) IMF,
\begin{eqnarray} M_1(x) = 0.19~x~[(1-x)^{3/4} +0.032(1-x)^{1/4} ]^{-1},
\end{eqnarray}
where $x$ is a uniformly distributed random number.  This
distribution flattens out toward lower masses, in contrast
with a Salpeter-type power-law IMF (1955). We considered primary
stars whose mass is in the range of $0.8 < M_1 < 8~M_\odot$.
Next, the mass of the secondary, $M_2$, is chosen from the
probability distribution, $f(q) = 5/4~q^{1/4}$, where $q
\equiv M_2/M_1$ (Abt \& Levy 1978; but also see Halbwachs 1987).
This mass ratio distribution is, at best, poorly known empirically.
Our adopted distribution has the property
that the mass of the secondary is correlated with the
mass of the primary, but is not strongly peaked toward $q = 1$.
We find that our results are not very sensitive to the choice
of $f(q)$, unless an extreme is adopted such as the assumption that the
two masses are to be chosen completely independently of one another
(see, e.g., Rappaport, Di Stefano, \& Smith 1994; hereafter
RDS - their Table 2 and Fig. 4).  Secondary masses as small as
$0.09~M_\odot$ are chosen (we wanted to ensure that only stars
with masses clearly above the minimum main-sequence mass are included).
To choose an initial orbital period, a distribution that is uniform
in log$(P)$ over the period range of $1$ day to $10^6$ years is
used (see, e.g., Abt \& Levy 1978; Duquennoy \& Mayor 1991).
Since we consider only circular orbits the adopted orbital period
distribution more properly pertains to the tidally circularized orbits
than to the initial orbits of the primordial binaries.

After the masses and orbital period are chosen, the orbital separation is
calculated using Kepler's 3rd law. We utilize an analytic expression
for the relation among the core mass, the radius, and the total mass of
the primary to estimate the mass of the degenerate core, $M_{WD}$,
when the primary fills its Roche lobe.  The expression we used for
this purpose (see RDS) was designed to reproduce the features of
Figure III.2 of Politano (1988) and Figure 1 of de Kool
(1992), except that the core-mass radius relation for stars
with mass $\la 2~M_\odot$ was renormalized to match the fitting
formula of Eggleton (2000; see eq. [4] of Joss, Rappaport, \& Lewis 1987).
Mass loss via a stellar wind prior to the start of the first mass-transfer
phase was computed via an analytic expression derived by M. Politano
(1999, private communication).  In practice, the inclusion of this wind mass
loss does not significantly affect the results.

In order to select only systems which undergo a common
envelope phase we require that the radius of the Roche
lobe of the primary be larger than the radius of a star
of mass $M_1$ at the base of the giant branch
(see, e.g., Paczy\'nski 1965; Webbink 1979, 1985, 1992,
de Kool 1992, and references therein).  This ensures that
unstable mass transfer will occur on a timescale that is
substantially shorter than a thermal time, and should lead to a
common envelope phase.  Once mass transfer from the primary to
the secondary commences, we assume that a common envelope phase
occurs and compute the final spiral-in separation based on simple
energetic considerations (see, e.g., Taam, Bodenheimer, \&
Ostriker 1978; Meyer \& Meyer-Hofmeister 1979; Livio \& Soker 1988;
Webbink 1992; RDS; Taam \& Sandquist 1998).  The expression
we use for determining $a_f$, the final orbital separation after
spiral-in, is given by:
\begin{eqnarray}
\frac{\epsilon GM_{2}}{2} \left(\frac{M_{core}}{a_f} - \frac{M_1}{a_i}\right)
= \frac{GM_{env}(M_{env}+3M_{core})}{R_1}
\end{eqnarray}
where $M_{core}\equiv M_{WD}$ and $M_{env}$ are the core and envelope
masses of the primary, $R_1$ is the radius of the primary, $a_i$ is the
initial orbital separation, and $\epsilon$ is the energy efficiency
factor for ejecting the envelope.  We take $\epsilon$ to have a
value of 1.0 in our standard model. The two terms in parentheses
on the right hand side of equation (9) represent the binding
energy of the envelope of the primary to itself and
to its core.  The dimensionless coefficients multiplying
each term were computed for an assumed polytropic envelope
structure with polytropic index $n = 3.5$ (RDS).  For other similar
values of $n$ the ratio of $\sim3:1$ between the two coefficients
is roughly the same.  We assume that the duration of the
spiral-in is sufficiently short ($< 10^4$ yr; see above
references) that the mass of the secondary does not change
significantly during the common envelope phase.

After the spiral-in episode, the separation, the white
dwarf mass, the secondary mass, and the corresponding Roche-lobe
radius of the secondary are known.  If at the end of the common
envelope phase the secondary would already be overfilling its
Roche lobe, then we eliminate the system.  (In most cases, this
circumstance would be expected to lead to a merger of the
secondary star with the degenerate core of the primary, which
presumably would result in the formation of a giant star.)
In practice, if the Roche lobe is larger than $\sim20~R_\odot$,
then neither magnetic braking nor gravitational radiation
would bring the system into Roche-lobe contact before
the secondary would evolve past the base of the giant branch.  We
can also eliminate these systems since the ensuing mass transfer
would either be dynamically unstable (see, e.g., Paczy\'nski 1967;
Kippenhahn, Kohl, \& Weigert 1967; Webbink 1979, 1992) or lead
to an even wider orbit; either case would not produce a CV of
the ordinary kind.

We typically start with $3-5 \times 10^6$ primordial binaries and
end up with $\sim15,000$ pre-CVs to evolve through the
mass-transfer phase with the bipolytrope evolution code described in
the next section.  The computational time for this first portion of the
calculations is negligibly short.

\subsection{Evolving the CVs Through Their Mass-Transfer Phase}

As mentioned earlier, the evolutionary tracks of CV systems
are calculated using a version of the code that
was first developed by RVJ (see also RJW)
to explore the effects of the parameterized Verbunt \& Zwaan (1981)
magnetic braking law on the evolutionary
properties of cataclysmic variables.  According to their algorithm, the
mass losing donor is approximated by a bipolytrope wherein the convective
envelope is represented by an n=3/2 polytrope and the radiative core by an
n=3 polytrope.  One of the advantages of this code is that it allows for
the rapid computation of a large number of evolutionary tracks and provides
a more physically intuitive interpretation of the results.  The original
version of the code has been modified substantially to allow
for improvements to the input physics, and to ensure that the conditions
near the surface (atmosphere) are more physically realistic.
A number of these changes have been discussed in previous papers.

The most significant of these modifications and updates are described
by Nelson, Rappaport, \& Joss (1986a, 1986b; 1993) who used a single polytrope
model to follow the evolution of fully convective low-mass stars and brown
dwarfs.  The results of the brown dwarf cooling evolutions and the
calculation of ZAMS models of low-mass stars are in excellent agreement
with those calculated using more sophisticated techniques (see, e.g.,
DNC; Burrows et al. 1993; Burrows et al. 1997; Baraffe et al.
1998, and references therein).  Specifically, coulombic corrections to the
pressure equation of state were incorporated and an updated version of the
Alexander, Johnson, \& Rypma (1983; Alexander [1989]) low-temperature,
radiative (surface) opacities was used.  The molecular hydrogen
partition function was also calculated more accurately.  Most importantly,
the specific entropy at the surface was matched directly to the specific
entropy in the interior; i.e.,
at the interface between the radiative core and the
convective envelope.\footnote
{A small entropy mismatch was introduced to correct for thin regions of
superadiabatic convection/radiative transport that exist beneath the
photosphere of the more massive stars in our mass range.  These corrections
depend on the assumed value of the mixing length parameter and
were chosen so as to provide the best possible representations of ZAMS stars.
They were largest for the $1.0~M_\odot$ model ($\sim 5\%$ of the specific
entropy), decreasing to zero for fully convective stars.}

In addition to these changes, the atmospheric pressure boundary condition
was modified so as to approximate more closely the scaled solar T-{\rm
$\tau$} (Krishna-Swamy 1972) relation.  The radiative surface opacities did
not include the effects of grain formation.  Since grains can only form in
the atmospheres of very low-temperature stars ($\leq$ 1500 K), this should
affect mostly the evolution of those CVs that have evolved
beyond the orbital period minimum. However, we have found that
the evolution of CVs through and beyond the period minimum,
is not particularly sensitive to this omission.

The overall result of all of these changes is that the theoretical
radius-mass relation for our ZAMS models with masses $\leq 1.0 M_\odot$
is now in substantial agreement with other theoretical calculations as well
as with observational studies of low-mass stars (see DNC). For similar
abundances
of hydrogen and for stars of approximately solar metallicity, we find that
the radii of our new models compared with other theoretical models (and the
DNC results) typically agree to within an rms error of $\sim 3\%$ ($M
\leq 1.0~M_\odot$).  Deviations among the theoretical models are greatest for
the higher mass stars due to uncertainties in the mixing length parameter
and the treatment of inefficient (superadiabatic) convection.  When
observations of double stars are considered, we believe that our ZAMS radii
are accurate to within $\sim 5 \%$. Our ZAMS models become fully convective
at a mass of $\sim 0.34~M_\odot$.  This is considerably smaller than the
value given in RVJ but agrees well with the DNC results (as well as
with newer generations of models).

Mass transfer in CVs is driven by angular momentum losses due to
gravitational radiation (Landau \& Lifshitz, 1962) and other systemic
angular momentum losses such as ``magnetic braking".
The magnetic braking law that we utilize is that of Verbunt \& Zwaan (1981)
and parameterized by RVJ.  The magnetic braking parameters were chosen so
as to best reproduce the observed period gap.  According to the
parameterization described in RVJ, we took $\gamma = 3$ and did {\it not}
adjust the multiplicative constant (defined here as $C_{MB}$) used in the
RVJ prescription. We also ``shut off" magnetic braking when the radiative
core had been reduced to less than 15\% of the mass of the
donor.  Magnetic braking is assumed to be greatly reduced as a result of the
restructuring of the magnetic field of the donor star when it becomes nearly
fully convective. This reduction in the angular momentum loss rate
gives the donor an
opportunity to shrink inside of its Roche lobe on a thermal timescale.
Further angular momentum losses due to gravitational
radiation cause mass transfer to recommence once the binary system is
brought back into a state of semi-detachment (see, e.g., RVJ; Spruit
\& Ritter 1983; Hameury et al. 1988a for a more detailed explanation).
As pointed out in several places in this work, the actual mechanism that
produces the bloating of the donor, and the means by which mass
transfer is interrupted, are not central to the conclusions drawn in this
paper. What is important in this regard is that the bloating be sufficiently
large as to produce the observed width of the period gap.  For our standard
evolutionary model the period gap covers the range of 2.1 $< P_{orb} <$
2.85 hr. According to Warner (1995), this synthetic gap approximates the
observed one very well.

We assume that mass and orbital angular momentum {\it are} lost
as a result of nova explosions on the surface of the white dwarf accretor.
For our standard model we assumed that all of the mass that is accreted
by the white dwarf is lost with the same specific angular momentum
as the white dwarf itself (see Schenker, Kolb \& Ritter 1992).  Given the
relatively low mass transfer rates, it is likely that the nova events are
extremely hydrodynamic, and thus it is unlikely that any of the accreted
mass actually contributes to increasing the mass of the white dwarf
(see, e.g., Prialnik \& Kovetz 1995, Starrfield 1998, and references therein).

After a potential cataclysmic variable system has been generated with the
population synthesis code, the two detached components are given the
opportunity to come into contact, via magnetic braking, within the age of
the Galaxy (minus the CV formation time). However, the
initial mass transfer may actually
be unstable, thereby leading to a common envelope phase (and the
ultimate demise of the binary system).  As derived by RJW, the
expression for the long-term mean mass transfer rate in a CV is
given by $|\dot M|/M$ = $N/D$, where the numerator, $N$,
contains the drivers of mass transfer, e.g., systemic angular momentum
losses, and the thermal expansion/contraction of the donor star (see
eq. [33] in RVJ).  The denominator is given by
\begin{eqnarray}
D = \left[ \left(\frac{5}{6} + \frac {\xi_{ad}}{2}\right) -
\frac {(1-\beta)q}{3(1+q)} - (1-\beta )\alpha (1 + q) - \beta q \right]
\end{eqnarray}
where $q \equiv M_{2}/M_{WD}$ (note that this is the inverse of the definition
used in RVJ), $\beta$ is the fraction of the mass lost by the donor star
that is ultimately retained by the white dwarf, $\alpha$ is the specific
angular momentum carried away by matter ejected from the binary system in
units of the binary angular momentum per unit reduced mass, and
$\xi_{ad}$ is the adiabatic index of the donor star, i.e.,
$[{\rm{dln(R)/dln(M)}}]_{ad}$. For our Standard Model (see Table 1), we
take $\beta = 0$ (i.e., all the mass accreted by the white dwarf is
eventually ejected in nova explosions\footnote{See Schenker et al. (1998)
for a justification as to why it is valid to approximate the ejection
of mass in a series of nova explosions with a constant value of
$\beta = 0$.}), and $\alpha ={M_2}^2/\left(M_2+M_{WD}\right)^2$.
With these definitions, the above equation reduces to
\begin{eqnarray}
D = \frac{5}{6} + \frac {\xi_{ad}}{2} - \frac {q~(1+3q)}{3(1+q)}
\end{eqnarray}
As discussed by RJW, stable mass transfer requires $N > 0$ and $D > 0$.  As an
example, consider donor stars with $M_2 < 0.3~M_\odot$ and $\xi_{ad}$ = -1/3.
In this case, stability (based on eq. [11]) requires that $M_{2} < M_{WD}$.
This allows for considerably larger values of $M_2$ than the more conventional
limit for conservative transfer where $M_{2} < 2/3~M_{WD}$ is required for
stable mass transfer (with low-mass unevolved donors).  Thus, the mass ratios
that appear in our population synthesis can often approach unity or exceed it.

\subsection{Generating the Population Synthesis Tracks}

We define a birth rate function, $BRF(t)$, for the progenitor primordial
binaries, where $t$ is the elapsed time between the formation of the Galaxy
and the birth of the primordial binary.  If a binary is born at time $t$, then
an additional time $\tau_{{\rm prim}}$ must elapse before the primary
evolves to the point where a common envelope phase may occur
(see \S 4.1).  We define this time ($t + \tau_{{\rm prim}}$) to be
the birth time of the incipient CV.  The resultant zero-age CV is then
evolved in the binary evolution code for a total time $t_{max}$ =
($10^{10}$  - $\tau_{{\rm prim}}$) yr, which is the maximum time
any CV that is descended from a similar primordial binary could evolve
before the current epoch.   (The binary evolution code starts with the
white dwarf and companion star as they emerge from the common
envelope, so the elapsed time, $t_{ev}$, includes the interval before
the donor star fills its Roche lobe.)   At each step in the evolution
code, specified by time $t_{ev}$ (with respect to the first time step in
the code), we sum in discrete binned arrays for various combinations
of $P_{orb}$, $M_{2}$, $M_{{\rm{WD}}}$, $q$, $\dot M$,
$T_{eff}$, and $L_2$, the following quantity, $\Delta Q$:
\begin{eqnarray}
\Delta Q = \frac{\Delta t\times BRF(10^{10} - \tau_{{\rm prim}} -
t_{ev})}{N}  .
\end{eqnarray}
In this expression, the argument of $BRF$ is the time that the primordial
binary was born with respect to the formation of the Galaxy, $\Delta t$
is the time interval for that particular step in the evolution run, and $N$
is the total number of systems that are selected to start the population
synthesis run.  For all of the population synthesis runs in this study, the
$BRF$ was taken to be constant in time.  Even though we have adopted a
constant
stellar birth rate per unit time, the method we use for generating the CV
population at the current epoch is completely general (see also Kolb 1993).

The net result of this procedure is that the sum of the $\Delta Q$s at the
end of the population synthesis run, in any particular bin, represents the
number of CVs at the current epoch with that particular parameter value.

\section{Population Synthesis Results}

The computed population of current-epoch CVs as generated by the above
techniques is displayed as a sequence of color images in Figures 3 through
7.  In Figure 3 we show the model CV population in the $\dot M-P_{orb}$ plane
for our standard model (cf. Fig. 2a). The image is generated in such a way
that the color reflects the logarithm of the number of current-epoch CVs at
a particular location in the $\dot M-P_{orb}$ plane. In each of the images the
color scale is located on the right side.  The image in Fig. 3 is comprised
of 100 pixels per hour interval in $P_{orb}$, and 100 pixels per decade in
$\dot M$.

The most noteworthy features in Figure 3 include the
distinct groups of systems located above and below the period gap. Note the
substantial difference in $\dot M$ for systems above and below the period gap;
for the latter systems only gravitational radiation losses drive mass
transfer. The minimum orbital period ($P_{min}\sim$65 min) is also clearly
evident, as are systems that have evolved well past the minimum period back
up to values of $P_{orb} \sim 2$ hr. It has been proposed that these
latter systems may be related to the so-called TOADs (``Tremendous
Outburst Amplitude Dwarf Novae"; see, e.g., Howell et al. 1995, HRP).
In the systems above the gap, there is a central band of evolutionary tracks
(blue and green) where a typical CV is most likely to be found at a
particular point in time during its evolution.  One also notices very short
lived episodes (red and yellow) of high mass transfer rates.  These occur
for individual systems as the donor star first fills its Roche lobe and
commences mass transfer, but before it can come into a quasi-steady state
of mass transfer (see discussion in \S 2).  The same type of behavior is
seen (green structure) for systems that have come into contact for the first
time below the period gap, i.e., with initially very low-mass donor stars.
The two main tracks (purple) evident in the systems below the period gap
are for He (lower) and CO (upper) white dwarfs, respectively.  We also
call attention to the small vertical (blue) feature at $P_{orb} \sim 2$ hr.
This may be related to the statistically significant larger number of CVs
with periods in the range of $110-120$ minutes (first pointed out by
Hameury et al. 1988b).  Finally, we note that there are systems found
within the period gap, though fewer per period interval than for systems
below the gap.  Systems found within the period gap are typically
ones which had initial donor masses of $\sim$0.22-0.34 M$_{\odot}$ and
commenced Roche-lobe overflow at orbital periods in the range of $\sim$2-3 hr
(see also Fig. 2 and its associated discussion).

Further, in regard to the $\dot M-P_{orb}$ plane shown in Figure 3, we
point out that for systems above the period gap, the width of the distribution
in $\dot M$ at any fixed value of $P_{orb}$ is only about a factor of $\sim$2
(we define the ``width" as containing $\sim 80\%$ of the systems).
This is in contrast with the {\it observed} spread in $\dot M$ for
CVs which is closer to an order of magnitude (see, e.g.,
Patterson 1984; Warner 1995). One cause of this spread may be the inherent
uncertainty in translating observed parameters into accurate estimates of $\dot
M$. Additionally, some of this discrepancy
might be resolved by the inclusion of the effects
of nova explosions in CVs which, on a quasi-regular basis, slightly
increase (or perhaps even decrease) the orbital separation (e.g., by
$\delta a/a \sim10^{-4}$) which is sufficient to change $\dot M$ appreciably
for some interval of time (see, Shara et al. 1986; Schenker et al. 1998;
Kolb et al. 2000).  However, we note that Schenker et al.
(1998) showed that, except for extreme model parameters, the occurrence of
the nova explosions generally does not substantially affect the overall
secular evolution of the CVs.  Therefore the main results and conclusions
presented in this work should be robust even without the inclusion of
{\it orbital} perturbations due to nova explosions (we do, in fact, take
into account the mass and angular momentum lost in such events).

The population of current-epoch CVs in the $R_2-P_{orb}$ plane for our
standard model is shown in Figure 4. The shape traced out in this figure
represents a statistical ensemble of the type of evolutions graphed in
Figs. 2c and 2f.  The usual features of the ``upper branch" of systems
above the period gap, systems in the ``lower branch" below the gap, and the
minimum orbital period are all represented in this figure.  Again, as in
Fig. 3, we see that some systems are formed within the period gap.  It is
difficult from this image to judge quantitatively how many systems are in
the gap, versus the density of points just below the gap.  This is
quantified later in this section (see Fig. 9).  Note that an
extrapolation of the ``upper branch" to
shorter orbital periods would undershoot the ``lower branch", on which the
stars are close to thermal equilibrium.  As discussed in \S 2, this
undershooting actually  (counterintuitively) results from the thermal
{\it bloating} of the donor star when it has a higher mass loss rate which is
driven by magnetic braking.  In particular, see equation (7) where we show
that $R_2$ scales as $f^{-0.65}$, where $f$ is the bloating factor.
The low-density features (yellow) just above the main tracks through the
``upper branch" are systems that have just come into Roche-lobe contact
for the first time and have not yet established a quasi-steady state of
mass transfer.  The blue-green ``thumb" feature just below the main track
of the ``upper branch" near the top edge of the period gap, represents
systems with He white dwarfs and donor stars of comparable mass that have
just come into
contact. Their mass transfer rates are higher than normal for these orbital
periods; thus the bloating factors for these donor stars are significantly
larger than for systems on the main track (cf. Fig. 2f).

Perhaps the most dramatic demonstration of the effects of thermal bloating
of the donor star can be seen in Figure 5 which shows the population of
current-epoch CVs in the $M_2-P_{orb}$ plane for our standard model.
The shape traced out in this figure represents a statistical ensemble of
the type of evolutions graphed in Figs. 2b and 2e.  All of the features
that appear in the $R_2-P_{orb}$ image (Fig. 4), also appear in this
$M_2-P_{orb}$ image,
except in a more exaggerated form.  This is a direct result of the simple
scaling argument summarized in equation (6) in \S 2, which indicates
that the bloating effect on the masses just above the period gap scales as
$M_{2} \propto f^{-1.95}$.  A casual inspection of Figure 5 shows that the
masses of the donor stars in CVs with periods just above the period gap are
fully $\sim40\%$ lower than would be expected if their radius-mass relation
followed that of main-sequence stars.  It is this effect that we propose be
used to discriminate between the currently held explanation for the period
gap and alternate scenarios.  We return to a quantitative discussion of
this issue in the next section.

Lastly in regard to the color images of the $R_{2}-P_{orb}$ and
$M_{2}-P_{orb}$ planes (Figs. 4 and 5), we comment on the relatively
large spreads in $R_2$ and $M_2$ for systems above the period gap
in contrast with those below the gap.  As we showed in equations (6)
and (7), for a fixed value of the bloating parameter $f$, both $M_2$ and
$R_2$ are unique functions of the orbital period (which would imply
narrow tracks).  For systems well above the period
gap, the Kelvin timescale, $\tau_{KH}$, is shorter than the mass-loss
timescale, $\tau_{\dot M} \equiv M/\dot M$, but, as the orbital period
decreases and approaches the period gap, the two timescales become
more comparable.  Thus, as discussed in \S2, the donor star
must become ever more bloated so as to establish a luminosity deficit,
which in turn enables the donor to contract inside its ever shrinking
Roche lobe.  Additionally, the adiabatic stellar index is changing
from positive to negative, and this tends to make the star expand
even further as it loses mass (see also Beuermann et al. 1998).
These two effects lead to the bloating behavior that is seen in Figs. 4
and 5.  The actual amount of bloating depends upon the absolute
values of the two constituent masses as well as on the thermal
history of the donor; therefore, it is to be expected that $f$ may
vary from one donor star to another.  As a result, we not only
see enhanced bloating as systems approach the period gap,
but a relatively wider and wider spread in the values of
$M_2$ and $R_2$ for these systems, especially
for $P_{orb}$ in the range of 3-5 hr.  By contrast, for systems just
below the period gap, $\tau_{\dot M}$ increases abruptly  - by about an
order of magnitude - because the mass transfer is then driven only by
gravitational radiation losses (at least according to our model), and
therefore the donor stars can remain much closer to thermal equilibrium.
This allows the systems right below the gap to establish a nearly
main-sequence radius-mass relation (i.e., $f\simeq1$), thereby leading
to a relatively narrow set of evolution tracks.  However,
as the secondary's mass approaches the minimum main-sequence mass
(before the orbital period minimum), $\tau_{KH}$ becomes very long
(due to a sharp decrease in the secondary's nuclear luminosity), thereby
causing $\tau_{KH}$ and $\tau_{\dot M}$ to again become approximately
equal. Thus, the width of the tracks broadens somewhat near the orbital
period minimum.  For systems beyond the orbital period minimum, the interiors
become increasingly electron degenerate.  This leads to a nearly unique
mass-radius relationship $R_2 \propto {M_2}^{-1/3}$ which, in
turn, leads to entirely different $R_{2}(P_{orb})$ and $M_{2}(P_{orb})$
relations than are given by equations (6) and (7).  Nonetheless, they are
unique relations (easily derivable from eqs. [4] and [5]) which also lead
to a very narrow set of tracks in Figs. 4 and 5.

The distribution of expected mass ratios, $q$, in CVs at the current epoch
is shown as a function of orbital period in Figure 6.  At any given orbital
period the range of $q$ values is considerably broader than the
distribution of values of $R_2$ or $M_2$ as can be seen by comparison with
Figs. 4 and 5.  The reason for this is straightforward.  Equations (6) and
(7) indicate that, as long as the bloating factor $f$ depends largely on
the orbital period of a CV, then both the radius and mass are nearly unique
functions of the orbital period.  Thus, the much broader distribution of
$q$ in Fig. 6 is due largely to the substantial range of masses that the white
dwarf may have, which is much less constrained by the orbital period
than is $M_2$.  For both the systems above and below the period
gap, the upper set of tracks corresponds to He white dwarfs, while the lower
tracks are for CO white dwarfs.  The period gap is especially conspicuous
in this figure, especially for systems with CO white dwarfs.   Note that
some of the mass ratios extend up to values of unity and, in some cases,
above unity.  The stability of mass transfer in these systems was discussed
in \S 4.2  [see equation (11)].

The evolution of our model population of CVs in the $T_{eff}-P_{orb}$
and luminosity$-P_{orb}$ planes is shown in Figure 7.  The left panel
displays the effective temperature, $T_{eff}$ of the donor star, while
the right panel has a superposition of the optical (bolometric) luminosity,
$L_{opt}$ and nuclear luminosity $L_{nuc}$.  Where the two sets of
luminosity tracks cross (e.g., near $P_{orb}$ = 1 hr) or overlap (e.g.,
to a minor extent between 3 and 6 hrs), the default is to display $L_{opt}$.
With regard to the $T_{eff}$ curves, we first note that the absolute
temperature scale for our main sequence stars (based on our
bipolytrope code) is somewhat shifted from that
produced with more sophisticated codes, e.g., our bipolytrope
main-sequence models are $\sim$200 K higher than the DNC models over
the mass range of $0.85-0.1 M_\odot$.  However, aside from this small
quantitative difference we are
confident that the overall qualitative trends and shapes of these tracks
are highly indicative of the behavior and properties of the donor through
its evolutionary history.  Note that for values of $P_{orb}$ below the gap
as well as above $\sim$5 hr, the $T_{eff}$ tracks are quite narrow, in analogy
with the tracks in the $M_{2}-P_{orb}$ and $R_{2}-P_{orb}$ planes,
since the donor stars are typically quite close to thermal equilibrium.  By
contrast, within the period range of 3-5 hr, $T_{eff}$ of the donors is
systematically lowered by up to 250 K compared with $T_{eff}$ of main
sequence stars at the same $P_{orb}$. This lower temperature amounts
to a change to a later spectral type (at a given $P_{orb}$) of $\sim2-4$
in decimal subclass. Additionally, we can see from Fig. 7 that, over this
same period range, the use of temperature (or spectral type) to determine
the mass of the secondary star would require very precise measurements,
since the expected $T_{eff}-P_{orb}$ distribution is relatively flat. We
draw two conclusions from this figure: (1) an observationally produced
$T_{eff}-P_{orb}$ or spectral type-$P_{orb}$ relation for CVs should indeed
yield a fairly simple shape (especially when smoothed out by uncertainties
in the measurements), and (ii) the use of $T_{eff}$ or spectral type in the
3-5 hr period range will not yield reliable indications of the mass of the
donor star.

Recently, Smith \& Dhillon (1988)\footnote{The sample
used in Smith \& Dhillon consisted of what are believed to be 55 reliable
spectral types and 14 reliable secondary star masses. All systems in their
sample have $P_{orb}>$ 90 min and $V_{min}$ brighter than $\sim$17th
magnitude, thus Smith \& Dhillon's conclusions about finding no evidence
for post-period minimum systems or very low-mass brown dwarf-like
secondaries cannot be drawn from the sample they used.} published
results which took a critical look at the relation between orbital period,
spectral type, and secondary mass based on the best observational
data available in the literature. They presented a relatively smooth spectral
type-orbital period relation but concluded that one cannot reliably estimate
$M_2$ in any given CV based solely on its spectral type. Our theoretical
results are quite consistent with their conclusion.

The image in the right panel of Figure 7 displays a superposition of the
evolutionary tracks for $L_{opt}$ and $L_{nuc}$ as functions of $P_{orb}$.
For systems above the gap, the highest luminosity track corresponds to
$L_{opt}$, while the two prominent lower (green) tracks are for $L_{nuc}$
and are related to the corresponding features in the $M_{2}-P_{orb}$ image.
These two lower luminosity tracks are for systems with CO (upper) and He
(lower) white dwarf accretors (see the discussion of Fig. 5). The
large luminosity deficit in the $P_{orb}$ range of 3-5 hr, already
discussed in \S2, shows up quite dramatically in this Fig. 7.  The group of
systems with the highest luminosity deficit (with He white dwarf accretors)
has the largest values of $\dot M$ and the donors are the most out of
thermal equilibrium (largest bloating factor).  In spite of the relatively
low values of $L_{nuc}$ in this period range, the bolometric luminosity
$L_{opt}$ is depressed only modestly (e.g., by factors of $\la $ 2) over
main-sequence stars at the same orbital period.  For systems below the
period gap, both luminosities fall off dramatically, especially for donor
masses below $\sim0.05-0.08~M_\odot$, where the donors are already below
the hydrogen-burning main sequence, and are cooling toward their ultimate
degenerate state.  The higher track for all points below the period gap
corresponds to $L_{opt}$, the lower one to $L_{nuc}$. While it is
formally true that the $L_{opt}$ and $L_{nuc}$ tracks ``cross" at
$\sim 10^{-4} L_\odot$, the two luminosities are never equal in
this part of the diagram; they reach the crossing point at
very different times.

Finally, with regard to the color image representations of CV populations in
parameter space, we note that in Figs. 3 through 7, the color represents
the logarithm of the numbers of systems expected at the current epoch.  As
can be seen in any of these figures (but especially Figs. 3 and 6), the
number of systems below the period gap outweighs the number
above the period gap by a
large margin (by about 100:1; see the more quantitative discussion below;
see also de Kool 1992, Kolb 1993). However, due to observational
selection effects, the systems with the shorter orbital periods, lower
values of $\dot M$, and generally longer intervals between dwarf-nova
outbursts, are more difficult to discover.  The exact factors that go
into the observational selection effects are complex, especially since some
CVs are discovered via their dwarf-nova outbursts, others (e.g., longer
period CVs) by their blue colors or flickering behavior, and still others
by their nova outbursts.  Some of these issues are discussed in RJW and
Kolb (1993). For purposes of the present work we will indicate only
qualitatively how the numbers of observationally known CVs might be
expected to be distributed by orbital period.  We adopt two crude
detectability factors which scale simply as $\dot M^{3/2}$ and as $\dot M$.
The first of these is appropriate to steady-state accretion luminosities, that
are proportional to $\dot M$ in the optical bandpass, which give rise to a
bolometrically flux-limited detectability proportional to $\dot M^{3/2}$
(analogous to the 3/2 slope of a log(N)-log(S) curve for isotropically
distributed sources).  The other scaling, appropriate to the case where
the flux in the optical bandpass is proportional to $\dot M^{2/3}$ (see
Lynden-Bell \& Pringle 1974; RJW; Webbink et al. 1987),
leads to a flux-limited detectability proportional to $\dot M$.  Other
factors leading to the discovery of CVs, beyond the simple consideration
of flux limited samples, in particular the detection of dwarf nova outbursts,
would undoubtedly substantially modify the rudimentary dependences on
$\dot M$ that we use here for purposes of illustration.
In Figure 8 we redisplay Fig. 3, but this time rescaled by a factor of $\dot
M$. It is clear from a casual inspection of Fig. 8 that the number of
``detectable" systems above the period gap is now at least as great as for
those below the gap.  The actual quantitative values for this simple
scaling are presented below. We again caution, however, that either
an  $\dot M^{3/2}$ or $\dot M$ scaling is oversimplified.

The color images of parameter space shown in Figures 3 -- 7 can be
displayed in a somewhat more quantitative fashion by projecting the
numbers of systems onto the various axes and plotting the results as simple
histograms.  For example, the data used to produce any of the images, can
be projected onto the $P_{orb}$ axis to yield the orbital period
distribution.  The results are shown in Figure 9.  The solid histogram in
Fig. 9a is the distribution of CVs at the current epoch in the entire
Galaxy for our standard model (see Table 1).  The stellar birthrate
function and IMF in our Standard Model [eq. (8)] are normalized in
such a way that there are $\sim 0.6$ stars born in the Galaxy per year with a
mass $>0.8~M_\odot$, just above the threshold for producing a remnant white
dwarf by the current epoch.  Thus, the ``absolute values" of the numbers
plotted in Fig. 9 can be appropriately scaled up or down for either lower or
higher assumed birthrates.

If the numbers of CV systems are scaled by the types of ``observability
factors" discussed above, before the histogram is produced, the results are
the dashed and dotted histograms superposed in Figure 9a.  As discussed
above, in conjunction with Fig. 8, this qualitatively helps to explain the
relative numbers of CVs observed above the period gap compared with
the number observed below (see especially the dotted histogram).
Inspection of the compilation of CVs with known
orbital periods given in Warner (1995) reveals that our histogram
shown in Fig. 9a with the $\dot M$ scaling provides qualitative agreement
with current observational results, especially considering the many
observational selection effects that exist (e.g., magnitude-limited color
surveys, large-amplitude but infrequent outbursts compared with
semi-periodic lower amplitude outbursts, discovery in X-ray surveys, etc.).
The distributions of orbital period shown in Figure 9b are for systems that
have not yet evolved to the minimum orbital period (solid curve), and
systems that have evolved beyond the period minimum (dashed curve) - no
scaling in $\dot M$ has been applied here.

The distributions of white dwarf masses and donor masses at the current
epoch are shown in Figure 10 (left and right panels respectively) for four
different ranges of orbital period. The He and CO white dwarfs are
easy to distinguish by mass.  Note that for systems with $P_{orb} > 4$ hr,
which typically have donor stars with masses $> 0.4~M_\odot$, there
are few He white dwarfs, since the mass transfer would tend to be unstable.
The distributions of donor star masses show a steady
trend toward higher masses at the longer periods, as expected.  This
results qualitatively from the fact that the larger orbital periods require
less dense, and therefore usually more massive, stars.  Note that the
distributions shown in this figure are not produced with sufficient
resolution in $P_{orb}$ to allow one to make quantitative predictions as to
what mass donors are needed to validate the basic paradigm for the period
gap.  Such information may be found, however, in Figs. 5 and 12, and Table
2.

The distribution of mass ratios $q$ ($\equiv M_{2}$/$M_{WD}$)
is shown in Figure 11 for two different orbital period ranges.
Attempts to determine $q$ observationally can be made
from, for example, superhump period analysis or spectroscopic analysis.
The observational distribution for $q$ in short period ($P_{orb} <2$ hr)
CVs has recently been compiled (Mennickent et al. 1999), and is seen
to show an approximate Gaussian distribution
with $<q>$ = 0.14. However, observational selection effects allow few
CVs with small $q$ to be discovered due to their intrinsic faintness.
Our results (Fig. 11 - top panel) show that the actual distribution
should not drop off at $q$ values lower then 0.14, but rather should
peak at values of $q$ = 0.05-0.1, with an overall distribution that is
clearly non-Gaussian. Discovery and observation
of additional faint (short period) CVs are needed in order to confirm
this theoretical prediction.

\section{Test of the Basic Paradigm}

The rapid rate of mass loss for donor stars in CVs just above the
period gap should lead to significant thermal bloating of the donor.  Thus,
in the conventional paradigm for the formation of the period gap, this mass
loss rate is abruptly decreased at orbital periods near 3 hr and the donor
star shrinks inside its Roche lobe (see \S 2), leading to the
cessation of mass transfer.  Specific choices of the parameters utilized in
any such evolutionary model change the bloating factor quantitatively, but
do not change the overall evolution qualitatively.  To demonstrate this, we
show in Figure 12 CVs at the current epoch in the $M_2-P_{orb}$ plane for
four different sets of model parameters (see Table 1).  Panel (A) is for
our Standard Model, while the other panels are for models where
(B) the proportionality constant in the magnetic braking formula was
reduced by a factor of 2 ($C_{MB} = 1/2$),  (C) the specific angular
momentum carried away by mass lost from the system in nova explosions is
twice that of the white dwarf ($\alpha = 2~\alpha_{WD}$), and (D) all mass
transferred to the white dwarf is ultimately retained by the white dwarf
(i.e., $\beta = 1$; in this somewhat artificial model, white dwarfs
are allowed to exceed the Chandrasekhar Limit).

We see from a study of Figure 12 that the effects of thermal bloating on
the mass of the donor stars in CVs for orbital periods just above the gap
are qualitatively similar for all four models.  The actual factors by which
the masses are lower than would be inferred by making the assumption that
the donor has a main-sequence mass-radius relation range from $25 - 50\%$;
the exact range depends on which model parameters are chosen and whether
one includes the CVs with He white dwarfs where the mass transfer can be
only marginally stable.  To quantify the effect of thermal bloating on mass
determinations, we have carried out weighted least squares fits of
polynomials to each of the ``upper branches" shown in Fig. 12.  The results
are given in Table 2 which also includes the evaluation of the polynomial
fit at $P_{orb} = 3$ hr.  As we can see from Table 2, the effect of thermal
bloating on the inferred donor mass of CVs is quite significant, and
potentially testable for any of these models.

We note that, in general, the {\it spread} in values of $M_2$, at a given
$P_{orb}$, around the best-fit $M_{2}(P_{orb})$ curve is substantially
smaller than the mean deviation from an $M_{2}(P_{orb})$ curve based
on the assumption of a main-sequence mass-radius relation, especially in
the crucial period range of 3-5 hr.  A large part of this spread is due
to the different values of the mass of the accretor, $M_1$, with lower
values of $M_2$ corresponding to the lower values of $M_1$ (see also
the discussion in \S 5).  However, we do not attempt here to produce
fits of the more general form $M_{2}(P_{orb}, M_{1})$.  Such fits
are not straightforward to construct since, among other things, there is the
added complication of the existence of a minimum value for $M_{1}$ at any
$P_{orb}$ (due to issues of mass-transfer stability; see \S 4.2).  In any
case, the main effect to be confirmed observationally concerns the
substantially reduced values of $M_2$ just above the period gap (3-5 hr),
compared to what would be expected if the donor stars followed a
main-sequence mass-radius relation.  If sufficient numbers of high
quality mass determinations of the secondary stars can be made, and
this basic effect is confirmed, then a secondary goal would be to
look for a weak, but positive correlation between $M_2$ and $M_1$.

In Figure 13 we plot the polynomial fits that we made to the upper branches
in the $M_{2}-P_{orb}$ plane for the four different models.  For comparison
we show the $M_{2}-P_{orb}$ relation that would be obtained if the donor
star followed a main-sequence radius-mass relation (the one derived from
our bipolytrope code).  This set of curves shows quantitatively how
mass determinations based solely on $P_{orb}$ are
affected by the thermal bloating effect.  Note how the effect should go
from a maximum at $\sim 3$ hr to quite small at $P_{orb} \sim 5.5$ hr.

Finally, we point out that if, in fact, the period gap is in any way
related to a relaxation from thermal bloating, then the inferred effect on
mass determinations based on the orbital period must be approximately in
the range of $25 - 50\%$.  To demonstrate this, we note that in the basic
paradigm for producing the period gap, the system masses do not change from
the upper boundary of the gap ($P_{upper}$) to the lower boundary
($P_{lower}$), while the radius shrinks from its bloated state,
characterized by a bloating factor, $f$, to nearly its main-sequence radius
at the lower edge of the gap.  A simple application of Kepler's 3rd law for
the case of a Roche-lobe filling star (which is true at both the upper and
lower edges of the gap) shows that
\begin{eqnarray}
f = \left(\frac{P_{upper}}{P_{lower}}\right)^{2/3}    ,
\end{eqnarray}
where $f$ must range from $\sim1.2 - 1.3$, depending on whether the period gap
is taken to be 3/4 of an hour in width or 1 hour, respectively (we have assumed
that the gap is centered at 2.5 hr). From equation (6) we see that this value
of $f$ should reduce the inferred mass, at the top edge of the period gap,
by amounts ranging from $\sim 30\%-40\%$, in basic agreement
with our more detailed population synthesis study. (For a related discussion
see Beuermann et al. 1998.)

This type of discrepancy between the mass inferred for a secondary star,
based on the CV orbital period and the assumption that its radius is
that of a main-sequence star
has probably already led to a number of incorrect mass determinations
reported in the literature, particularly for systems with $P_{orb}$ between
3-5.5 hr. For example, at an orbital period of 3.2 hours, the mass assigned
to a CV secondary would be 0.35 M$_{\odot}$ while our calculations
show that it would actually be only 0.26 $\pm 0.02 M_{\odot}$, although
bloated in size. For a known or inferred mass ratio of say $q = 0.4$, we
would then calculate a white dwarf mass of 0.89~M$_{\odot}$, when the true
white dwarf mass is only 0.65~M$_{\odot}$. Thus, ignoring the bloating
effect in the secondary stars in CVs with orbital periods of 3-5 hr, can
lead to a significant overestimation of both component masses.

It is interesting to note here that the secondary stars which are
farthest from thermal equilibrium are those in systems with orbital periods
just above the period gap (see Figs 4, 5 \& 13). Observationally, this
orbital period region (3-4 hr) essentially contains only high mass transfer
rate, novalike (NL) types of CVs. The inferred high mass transfer rates for
these systems would then be expected, on theoretical grounds, to lead to a
large bloating of the secondary stars, and hence lower masses than might
otherwise be anticipated. Precise observational determinations of the
secondary star masses in NLs would allow a confirmation of this basic
effect which is, in fact, required if the period gap is to be explained by
the interrupted magnetic braking scenario.

Figure 13 provides our theoretical predictions for the most likely
mass of the secondary star at any given orbital period (see Table 2).
Observational determinations accurate to a few percent would be
needed in order to differentiate between the four models presented;
but, accuracies of only $\sim10\%$ will allow a test
of the bloating model in general, and the predicted deviation of the donor
star from the main sequence. This is a challenging observational project,
however, since the systems with orbital periods in the 3-5 hr range are ones in
which the secondary star is rarely directly observed. IR spectral studies
(e.g., Howell, et al. 2000; Mason, et al. 2000; Dhillon, et al. 1997) have
looked in detail for the secondary star in a number of CVs with only marginally
successful results. In these CVs, spectral identification of absorption
features due to the secondary star is difficult since the lines are
rotationally broadened and filled in by radiation from the accretion disk.
For the critical 3-5 hr period range, a signal-to-noise ratio of
$>$100 in the continuum will be needed to allow the atomic and molecular
features of the secondary to be observed against the
high background accretion-disk dominated continuum.  We
therefore advocate high signal-to-noise, orbital phase-resolved,
near- and mid-IR spectroscopic observations with large
ground-based telescopes (e.g., Gemini, Keck), and eventually with
SIRTF, of sources such as the brightest NLs and other CVs which have
$P_{orb}$ = 3-5 hr.

\section{Summary and Conclusions}

In this paper we briefly reviewed our current understanding of the
secular evolution of CVs through their mass transfer phase, including the
currently accepted model for the 2-3 hr ``period gap" in the orbital period
distribution.  The results of evolution calculations for a representative
sample of individual systems are presented, both as functions of time and of
orbital period.  A population synthesis code, that starts with some $3
\times 10^6$ primordial binaries, was then used to generate $\sim2 \times
10^4$ systems which evolve successfully through the CV phase of mass
transfer.  This allows for a more complete exploration of parameter space.
The results are displayed as probability densities in the $\dot M-P_{orb}$,
$M_{2}-P_{orb}$, $R_{2}-P_{orb}$ $q-P_{orb}$, and $T_{eff}/L_{2}-P_{orb}$
planes, for CVs at the current epoch. This method of displaying the
results can lead to considerable insight into the relationships among
the various system parameters. We find that for CVs with orbital periods
above 5.5 hr and below the period gap (but above the period minimum)
the secondary stars closely follow the main-sequence R-M relation
(cf. Beuermann et al. 1998). However, for those with $P_{orb}$ between
3-5.5 hr, the effect of bloating causes them to deviate substantially from
this same relation.

Among our more interesting results, we have shown that the donor star
masses in CVs with orbital periods just above the period gap should be as
much as $30-50\%$ lower than would be inferred on the assumption that the
donor stars obey a main-sequence radius-mass relation.  This conclusion is
only valid if the basic underlying cause of the period gap is thermal
bloating of the donor star for systems above the period gap (see \S \S
1-6).  On the basis of our results, we have proposed a direct observational
test of, in particular, the basic paradigm of the period gap and, more
generally, our overall understanding of the evolution of CVs.  This test
involves the challenging, but realistic, task of making relatively accurate
(e.g., $10\%$) determinations of the secondary masses in about a half dozen
CVs in the period range of 3-4 hr. If the masses are consistent
with the
assumption of a main-sequence radius-mass relation for the donor stars,
then the currently accepted explanation of the period gap cannot be
correct, and the very existence of the gap would pose a major conundrum.
If, on
the
other hand, the masses are mostly consistent with the lower values predicted
in this work, then a substantial part of our basic understanding of the
secular
evolution of CVs will be validated.

Previously, much observational attention in CV studies has been focused on
determinations of the white dwarf masses.  While this is clearly of great
interest, we hope with this work to stimulate more interest in the important
issue of determining the secondary masses.

This research was supported in part by NASA under ATP grants
GSFC-070 and NAG5-8500 (to S.B.H.), and NAG5-7479 and NAG5-4057 (to S.A.R.).
L.A.N. acknowledges the financial support of NSERC (Canada) and thanks CITA
and the University of Toronto for a Reinhardt Fellowship and for their
hospitality.  We thank M. Politano for a number of useful discussions
relating to this work. We are grateful to an anonymous referee who made
numerous helpful and insightful comments that led to significant
improvements in the paper.
We also thank D. MacCannell and G. Esquerdo for their technical assistance.

\begin{deluxetable}{lcccc}
\tablenum{1}
\tablecaption{Summary of Model Parameters}
\tablehead{
\colhead{Model}
 &\colhead{$\beta$\tablenotemark{a}}
 &\colhead{$\alpha$\tablenotemark{b}}
 &\colhead{$\gamma$\tablenotemark{c}}
 &\colhead{$C_{MB}$\tablenotemark{d}}
}
\startdata
A - Standard Model     & 0 & 1 & 3 & 1 \\
B - Reduced Magnetic Braking         & 0 & 1 & 3 & 1/2 \\
C - High Angular Momentum Losses     & 0 & 2 & 3 & 1 \\
D - Conservative Mass Transfer      & 1 & -- & 3 & 1 \\
\enddata
\tablenotetext{a} {Fraction of mass lost by the donor star that is transferred
to, and ultimately retained by, the white dwarf.}
\tablenotetext{b} {Specific angular momentum carried away in nova
explosions in units of the specific angular momentum of the white dwarf.}
\tablenotetext{c} {Magnetic braking parameter ``$\gamma$" as defined in RVJ.}
\tablenotetext{d} {Proportionality constant in the magnetic braking expression
used by RVJ, in units of their ``standard" value.}
\end{deluxetable}{}

\begin{deluxetable}{lccccc}
\tablenum{2}
\tablecaption{Summary of Polynomial Fits to $M_{2}-P_{orb}$
Relations\tablenotemark{a}}
\tablehead{
\colhead{Model\tablenotemark{b}}
  &\colhead{$c_0$}
  &\colhead{$c_1$}
  &\colhead{$c_2$}
  &\colhead{$c_3$}
  &\colhead{$M_{2}(P_{orb} = 3$ hr)\tablenotemark{c}}
}
\startdata
A	 & 0.005863 & --0.001251 & 0.02353  & 0.0      & 0.214 \\
B       & --0.4323  & 0.3294   & --0.04942 & 0.005028 & 0.247 \\
C       & --0.1829  & 0.1031   & 0.01041  & 0.0      & 0.220 \\
D      & --0.5280  & 0.3856   & --0.06261 & 0.006076 & 0.230 \\
Main-Sequence Donor    & -- -- & -- -- & -- -- & -- -- & 0.35 \\
\enddata
\tablenotetext{a}{For systems with $P_{orb} >$ 3 hr.  Fits are of the form:
$M_2 = c_0 + c_{1}P_{orb} + c_{2}P^2_{orb} + c_{3}P^3_{orb}$}.
\tablenotetext{b} {Models are defined in Table 1.}
\tablenotetext{c} {In units of M$_\odot$.}
\end{deluxetable}{}

\newpage

Fig. 1 -- Evolution with time of the mass transfer rate, $\dot M$, and orbital
period, $P_{orb}$, for several model cataclysmic variable systems.  Left
panel - the evolution of a single CV with initial masses ($M_2 = 0.9
~M_\odot$;  $M_{WD} = 1.1~M_\odot$).  This system first comes into
Roche-lobe contact at $P_{orb} = 6$ hr, evolves through the period gap, to
the minimum in $P_{orb}$, and back up to longer periods by $10^{10}$ yr.
Right panel - the evolutions of a selection of four other illustrative
initial binary constituent masses, $M_2$,$M_{WD}$ = 0.2,0.4 (solid),
0.35,0.35 (dotted), 0.3,0.6 (dashed), and 0.65,0.7 (long dashed),
all in units of $M_\odot$.

Fig. 2 -- Evolution with orbital period, $P_{orb}$, of the mass transfer
rate, $\dot M$, secondary mass, $M_2$, and secondary radius, $R_2$, for
several illustrative model cataclysmic variable systems.  The initial
masses for the systems whose evolutions are displayed in the left and right
sets of panels are the same as described in Fig. 1.

Fig. 3 -- Computed population of cataclysmic variables at the current epoch
in the $\dot M-P_{orb}$ plane for our Standard Model (see Table 1).  Here
$\dot M$ is the mass transfer rate, and $P_{orb}$ is the orbital period.  The
color represents the logarithm of the number of systems in a particular
$\dot M-P_{orb}$ cell, of which there are 100 per hour interval in $P_{orb}$
and 100 per decade in $\dot M$.  The color scale is given on the right side
of the
figure. We note that the scattered, isolated (red) points in the image {\it
below} the main tracks are minor numerical artifacts of the evolution code
that occasionally appear when the Roche lobe makes initial contact with the
atmosphere of the donor star.  One of these dots corresponds to only
$\sim0.1$ CVs in the entire Galaxy at the current epoch, and so is of no
significance.

Figs. 4 -- Computed population of cataclysmic variables at the current
epoch in the $R_{2}-P_{orb}$ plane for our Standard Model (see Table 1).
Here $R_2$ is the radius of the donor star.  The color represents the
logarithm of the number of systems in a particular $R_{2}-P_{orb}$ cell, of
which there are 100 per 0.1 $R_\odot$ and 100 per hour interval in $P_{orb}$.
The color scale is given on the right side of the figure.

Figs. 5 -- Computed population of cataclysmic variables at the current
epoch in the $M_{2}-P_{orb}$ plane for our Standard Model (see Table 1).
Here $M_2$ is the mass of the donor star.  The color represents the
logarithm of the number of systems in a particular $M_{2}-P_{orb}$ cell, of
which there are 100 per 0.1 $M_\odot$ and 100 per hour interval in $P_{orb}$.
The color scale is given on the right side of the figure.

Fig. 6 --  Computed population of cataclysmic variables at the current
epoch in the $q-P_{orb}$ plane for our Standard Model (see Table 1); $q
\equiv M_{2}/M_{WD}$.  The color represents the logarithm of the number of
systems in a particular $q-P_{orb}$ cell, of which there are 100 per $\Delta
q = 0.1$ and 100 per hour interval in $P_{orb}$.  The color scale is given on
the right side of the figure.

Fig. 7 -- Computed population of cataclysmic variables at the current epoch
in the {$T_{eff}$ -- $P_{orb}$} plane (left panel), and the {$Luminosity -
P_{orb}$} plane (right panel) for our Standard Model (see Table 1). We show
both the stellar luminosity (top curve) and the core nuclear luminosity
(lower distributions). The color represents the logarithm of the number of
systems in a particular {$L - P_{orb}$} or {$T_{eff} - P_{orb}$} cell of
which there are
100 per decade in $L$, 100 per 500K in $T_{eff}$, and 100 per
hour interval in $P_{orb}$. The color scale for both plots is given on the
right.

Fig. 8 -- Same as Figure 3, except that the population has been scaled by
$\dot M^{1}$ to crudely take into account observational selection effects.

Fig. 9 -- Computed orbital period distributions for cataclysmic variables
at the current epoch. Left panel - solid curve is the distribution for all
systems that appear in Figure 3; the dashed curve was produced by scaling
the contributions of each system evolved by $\dot M^{3/2}$ while the dotted
curve is for an $\dot M^{1}$ scaling (see text). The $\dot M^{3/2}$- and
$\dot M^{1}$-scaled
curves have been shifted vertically by arbitrary amounts
for ease in comparison. Right panel - solid
curve is for all systems in Fig. 3 which have not yet reached orbital
period minimum; dashed curve is for systems that have evolved past the
orbital period minimum.

Fig. 10 -- Computed distributions of the secondary (right panels) and white
dwarf masses (left panels) in cataclysmic variables at the current epoch.
The mass distributions are ordered according to the range of orbital
period. The dotted histogram (upper right) is for post-period minimum CVs and
has been arbitrarily divided by 1.5 for presentation purposes.

Fig. 11 -- Computed distribution of mass ratios in cataclysmic variables at
the current epoch.  The top panel is for systems with orbital periods in
the range of 1-3 hr (which includes all post period-gap systems), while the
bottom panel is for systems above the period gap.

Fig. 12 -- Same as Figure (5), except that in addition to the Standard
Model (A), the results for three other models are shown (see Tables 1 and
2):  (B) reduced magnetic braking constant; (C) specific angular momentum
lost with the ejected matter is twice that of the white dwarf; and (D)
conservative mass transfer and retention by the white dwarf.

Fig. 13 -- Secondary (donor) mass, $M_2$ as a function of orbital period.
The solid curve is based on the assumption that the donor star fills its
Roche lobe and has a radius-mass relation appropriate to stars on the main
sequence (i.e., eq. [2]) The main-sequence models
were generated with the same bipolytrope code that was used to carry out
the binary stellar evolution calculations and are discussed in the text.
The dashed curves are
polynomial fits to the $M_{2}-P_{orb}$ relations derived from the
population synthesis study shown in Figure 12.  The labels, A through D,
correspond to the four different panels in Figure 12.

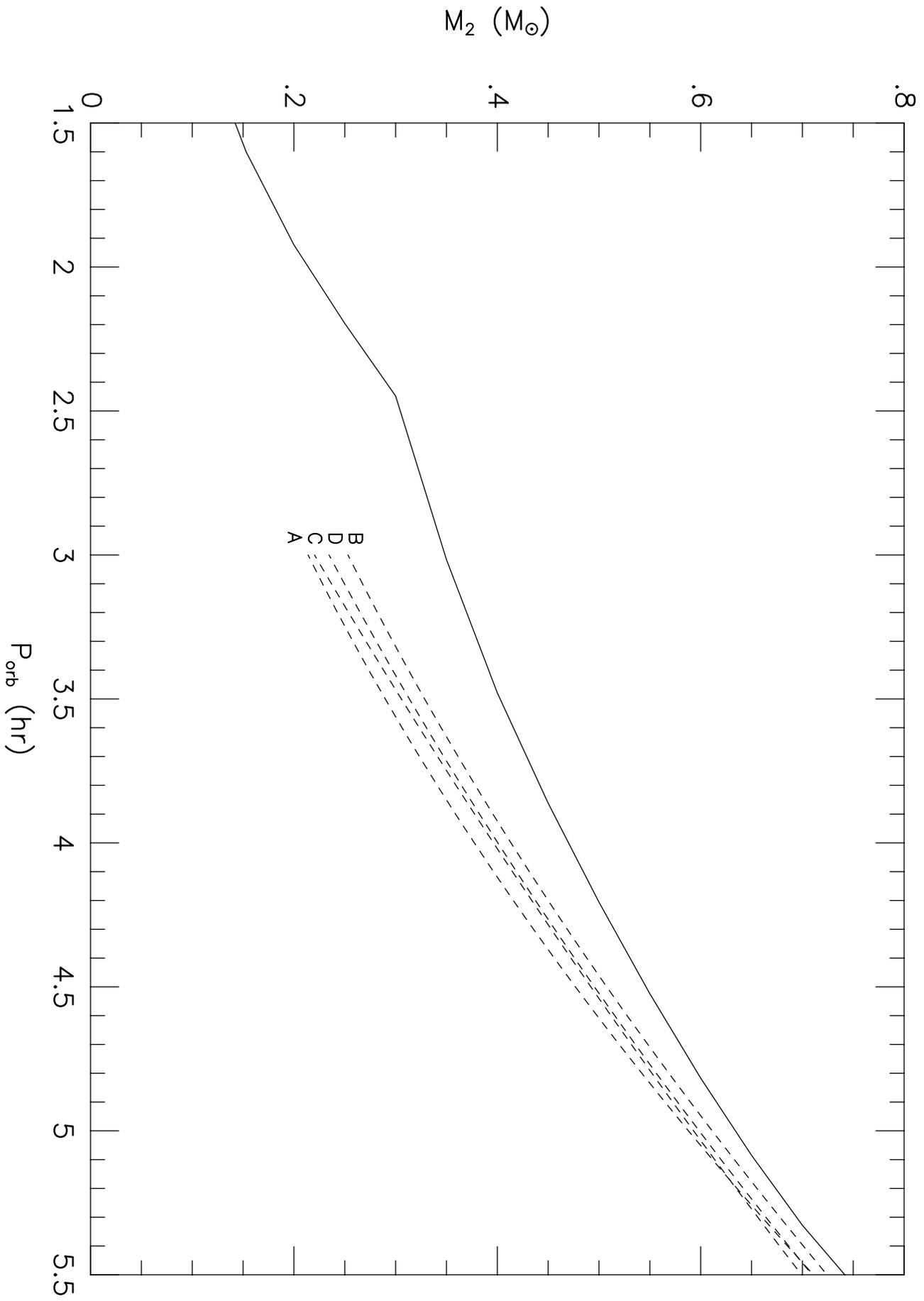
\end{document}